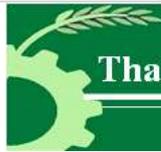

# On the Evaluation of Skill in Binary Forecast

*Thitithep Sitthiyot**
*Faculty of Commerce and Accountancy, Chulalongkorn University, Thailand*

*Kanyarat Holasut*
*Faculty of Engineering, Khon Kaen University, Thailand*

## Abstract

𝔄 good prediction is very important for scientific, economic, and administrative purposes. It is therefore necessary to know whether a predictor is skillful enough to predict the future. Given the increased reliance on predictions in various disciplines, a prediction skill index (PSI) is devised. Twenty-four numerical examples are used to demonstrate how the PSI method works. The results show that the PSI awards not only the same score for random prediction and always predicting the same value, but also nontrivial scores for correct prediction of rare or extreme events. Moreover, the PSI can distinguish the difference between the perfect forecast of rare or extreme events and that of random events by awarding different skill scores, while other conventional methods cannot and award the same score. The data on the growth of real gross domestic product forecast of the Bank of Thailand between 2000 and 2019 are also used to demonstrate how the PSI evaluates the skill of the forecaster in practice.

_________________________
* Corresponding author: Mahitaladhibesra Bld., 10th Fl., Phayathai Rd., Pathumwan, Bangkok 10330, Thailand. E-mail address: thitithep@cbs.chula.ac.th.

# 1. Introduction

The "Finley affair" is referred to the period from 1884 to 1893, where it is marked as the beginning of substantive, conceptual, and methodological developments and discussions in the field of prediction verification (Murphy, 1996). The event started in response to the work by Finley (1884), who published his experimental results of a binary prediction whether or not a tornado would occur. After more than a century, prediction verification has been undergoing significant developments, with more methods, scores, and techniques continually being invented and/or reinvented (Jolliffe & Stephenson, 2012). In the field of deterministic prediction skill verification for binary events, Hogan and Mason (2012) discuss eighteen of them (see also references therein). Jolliffe (2016) adds another called the Dice-coefficient commonly used in ecology. While the abundance of and the increased reliance on predictions in various professions provides exciting opportunities for innovative cross-disciplinary work in prediction verification (Jolliffe & Stephenson, 2012), Murphy (1996) notes that several deterministic prediction skill verification measures proposed in connection with the "Finley affair", namely, Gilbert skill score (GSS) (Gilbert, 1884), Heidke skill score (HSS) (Heidke, 1926) originally proposed by Doolittle (1888) who formulated the 2x2 version, Peirce skill score (PSS) (Peirce, 1884), and Clayton skill score (CSS) (Clayton, 1934) have stood the test of time and are still widely used nowadays.

There are various aspects of quality of prediction that constitute a good prediction, such as accuracy, skill, bias, and reliability, to name a few (Murphy, 1993). However, in this study, we focus only on skill in deterministic prediction of binary events. According to Wilks (2011), prediction skill is defined as the relative accuracy of a set of forecasts with respect to some set of reference forecasts whose common choices are average values of the predictand, values of the predictand in the previous time period, or random forecasts. Provided that it is essential to know if a predictor, either human or machine, has sufficient skill to predict the future, we view that this issue is of importance for scientific, economic, and administrative purposes as noted by Woodcock (1976) since deterministic prediction skill scores are widely used in various disciplines of science as a tool for decision-making. Examples include computer science – the performance of hybrid decision forest in short-term forecasting (Faisal, Monira, & Hirose, 2013); ecology – the measurement of the degree to which two different species are associated in nature (Dice, 1945); economics – the forecasting of binary outcomes (Lahiri & Yang, 2013); environment – the forecast of visibility degradation due to poor air quality (So, Teakles, Baik, Vingarzan, & Jones, 2018); finance – the prediction of stock market returns (Granger & Pesaran, 2000); hydrology – the development and verification of a real-time stochastic precipitation nowcasting system for urban hydrology (Foresti, Reyniers, Seed, & Delobbe, 2016); machine learning – the development and application of spatiotemporal machine learning/data mining techniques (McGovern, Gagne II, Williams, Brown, & Basara, 2014); the study of machine learning in space weather (Camporeale, 2019); medical and clinical studies – the mammogram screening (Briggs & Ruppert, 2005); the study of predictive models for decision-making against dengue haemorrhagic fever epidemics (Halide, 2009); meteorology – the behavior of verification measures with respect to random changes (Manzato & Jolliffe, 2017); seismology – earthquake forecasting and verification (Holliday, Rundle, & Turcotte, 2009); space weather – the study of an operational solar flare forecast (Kubo, Den, & Ishii, 2017).

Despite the fact that there are a number of methods available for assessing deterministic prediction skill, we view that we could contribute to the knowledge in the science of prediction verification by introducing a prediction skill index (PSI) as an alternative method for assessing skill in deterministic forecast of binary events. Our method utilizes all elements in the 2x2 contingency table, namely, *Hits, False alarms, Misses*, and *Correct rejections*. It awards the same score for random prediction and constantly predicting the same value. In addition, our PSI weighs correct predictions of rare or extreme events more strongly than correct predictions of more common events, and therefore could discourage distortion of prediction toward more common events in order to artificially inflate the score of prediction skill (Wilks, 2011). Given that the skill in prediction of rare or extreme events depends upon the current state of scientific knowledge used in prediction, we are well aware that not all rare or extreme events are difficult to predict. For example, solar eclipse is considered a rare event but not difficult to predict. Thus, in this study, rare or extreme events are referred to events commonly accepted as very difficult to forecast based on the current state of scientific knowledge used in forecasting such events. Examples of rare or extreme events are financial and economic crises, the success of start-up unicorns, droughts, floods, wildfires, earthquakes, wars, and pandemics. Given the importance of rare or extreme events as witnessed in our time and more to come in the future along with random events, as well as the impacts of these events on human societies and our planet, our PSI can distinguish the difference between the perfect forecast of rare or extreme events commonly accepted as very difficult to forecast and that of random events, while other conventional methods for deterministic prediction skill evaluation, namely, GSS, HSS, PSS, CSS, and Odds ratio skill score (ORSS) (Yule, 1900) cannot and award the same score. In this study, random events are defined as events where each of the binary outcomes (i.e. yes/no, head/tail, up/down, or 0/1) in every period has a 50:50 chance of occurring. An obvious example would be a fair coin-toss, which is generally considered extremely difficult to forecast given the current state of scientific knowledge. This should not be confused with binary events where each of the possible outcomes in every period has a 50:50 chance of happening but are not random, and hence not difficult to forecast based on the current state of scientific knowledge. A vivid example would be the pattern of day and night cycle.

This study is divided into 5 sections. Following the Introduction, Section 2 describes methods and data used in this study. Section 3 presents results and discussions. The difference between prediction skill and prediction accuracy is discussed and illustrated in Section 4. Finally, Section 5 provides conclusions and remarks.

## 2. Methods and Data

Let $a, b, c,$ and $d$ be the counts for *Hits*, *False alarms*, *Misses*, and *Correct rejections*, respectively. The total sample size, $n$, therefore, equals $a + b + c + d$. The 2x2 contingency table of two predicted and two observed events are shown as Table 1. Given the cell counts for *Hits* ($a$), *False alarms* ($b$), *Misses* ($c$), and *Correct rejections* ($d$), as well as the total sample size ($n$), the PSI score computed as a function of $a, b, c, d,$ and $n$ can be derived as follows.

Table 1: The 2x2 Contingency Table of Predicted and Observed Events.

|  | *Events observed* | | |
|---|---|---|---|
| *Events predicted* | Yes | No | Total |
| **Yes** | a (Hits) | b (False alarms) | $a + b$ |
| **No** | c (Misses) | d (Correct rejections) | $c + d$ |
| **Total** | $a + c$ | $b + d$ | $a + b + c + d = n$ |

Source: Authors' derivation.

We first begin by calculating joint and marginal probabilities based on the cell counts for *Hits* $(a)$, *False alarms* $(b)$, *Misses* $(c)$, and *Correct rejections* $(d)$ as shown in Table 1. The joint probabilities $\left(\frac{a}{n}, \frac{b}{n}, \frac{c}{n}, \text{and } \frac{d}{n}\right)$ and marginal probabilities $\left(\frac{a+b}{n}, \frac{c+d}{n}, \frac{a+c}{n}, \text{and } \frac{b+d}{n}\right)$ are shown in Table 2.

Table 2: The 2x2 Contingency Table Illustrating Joint and Marginal Probabilities.

|  | *Events observed* | | |
|---|---|---|---|
| *Events predicted* | Yes | No | Total |
| **Yes** | $\frac{a}{n}$ | $\frac{b}{n}$ | $\frac{a+b}{n}$ |
| **No** | $\frac{c}{n}$ | $\frac{d}{n}$ | $\frac{c+d}{n}$ |
| **Total** | $\frac{a+c}{n}$ | $\frac{b+d}{n}$ | 1 |

Source: Authors' derivation.

We then use the marginal probabilities to compute the expected values of a random prediction for *Hits* $(a)$, *False alarms* $(b)$, *Misses* $(c)$, and *Correct rejections* $(d)$. These values represent prediction that does not require any skill. They are shown in Table 3.

Table 3: The 2x2 Contingency Table Representing Expected Values for Prediction without Skill.

|  | *Events observed* | |
|---|---|---|
| *Events predicted* | Yes | No |
| **Yes** | $\left(\frac{a+b}{n}\right)\left(\frac{a+c}{n}\right)$ | $\left(\frac{a+b}{n}\right)\left(\frac{b+d}{n}\right)$ |
| **No** | $\left(\frac{a+c}{n}\right)\left(\frac{c+d}{n}\right)$ | $\left(\frac{b+d}{n}\right)\left(\frac{c+d}{n}\right)$ |

Source: Authors' derivation.

After we obtain the expected values of prediction without skill for *Hits* $(a)$, *False alarms* $(b)$, *Misses* $(c)$, and *Correct rejections* $(d)$, we can assess skill in prediction by calculating the difference between the joint probability and its corresponding expected value of a random prediction. In this study, we choose the expected value of a random prediction, which requires no skill, as our benchmark against which a prediction can be judged. Our prediction skill, therefore, is evaluated as the errors between the events predicted and the prediction of the events that does not require any skill. They are illustrated in Table 4.

Table 4: The 2x2 Contingency Table Illustrating Errors between Predicted Events and Expected Values for Prediction of Those Events without Skill.

| Events predicted | Events observed | |
|---|---|---|
| | Yes | No |
| Yes | $\frac{a}{n} - \left[\left(\frac{a+b}{n}\right)\left(\frac{a+c}{n}\right)\right]$ | $\frac{b}{n} - \left[\left(\frac{a+b}{n}\right)\left(\frac{b+d}{n}\right)\right]$ |
| No | $\frac{c}{n} - \left[\left(\frac{a+c}{n}\right)\left(\frac{c+d}{n}\right)\right]$ | $\frac{d}{n} - \left[\left(\frac{b+d}{n}\right)\left(\frac{c+d}{n}\right)\right]$ |

Source: Authors' derivation.

The next step is to normalize the errors between the events predicted and the expected values of prediction without skill by geometric mean of the expected value of a random prediction. The geometric means of the expected value of a random prediction for *Hits* $(a)$, *False alarms* $(b)$, *Misses* $(c)$, and *Correct rejections* $(d)$ are $\sqrt{\left(\frac{a+b}{n}\right)\left(\frac{a+c}{n}\right)}$, $\sqrt{\left(\frac{a+b}{n}\right)\left(\frac{b+d}{n}\right)}$, $\sqrt{\left(\frac{a+c}{n}\right)\left(\frac{c+d}{n}\right)}$, and $\sqrt{\left(\frac{b+d}{n}\right)\left(\frac{c+d}{n}\right)}$, respectively. The normalized errors between the predicted events and the expected values of prediction without skill are shown in Table 5.

Table 5: The 2x2 Contingency Table Illustrating Normalized Errors between Predicted Event and Expected Values of Prediction without Skill.

| Events predicted | Events observed | |
|---|---|---|
| | Yes | No |
| Yes | $\dfrac{\frac{a}{n} - \left[\left(\frac{a+b}{n}\right)\left(\frac{a+c}{n}\right)\right]}{\sqrt{\left(\frac{a+b}{n}\right)\left(\frac{a+c}{n}\right)}}$ | $\dfrac{\frac{b}{n} - \left[\left(\frac{a+b}{n}\right)\left(\frac{b+d}{n}\right)\right]}{\sqrt{\left(\frac{a+b}{n}\right)\left(\frac{b+d}{n}\right)}}$ |
| No | $\dfrac{\frac{c}{n} - \left[\left(\frac{a+c}{n}\right)\left(\frac{c+d}{n}\right)\right]}{\sqrt{\left(\frac{a+c}{n}\right)\left(\frac{c+d}{n}\right)}}$ | $\dfrac{\frac{d}{n} - \left[\left(\frac{b+d}{n}\right)\left(\frac{c+d}{n}\right)\right]}{\sqrt{\left(\frac{b+d}{n}\right)\left(\frac{c+d}{n}\right)}}$ |

Source: Authors' derivation.

It should be carefully noted, however, that if one or more of the contingency table cell counts is equal to zero, which would result in the PSI score to be undefined because there is no prediction or no observed event or both, this would make the denominator to be zero, or both the numerator and the denominator are equal to zero. If one of these is the case, Hogan and Mason (2012) recommend that the practical solution is to define them to take the no-skill value of 0.

To finish, the PSI can be computed as the sum of the normalized errors between the predicted events and the expected values of prediction without skill for correct predictions (*Hits* ($a$) and *Correct rejections* ($d$)) minus the sum of the normalized errors between the predicted events and the expected values of prediction without skill for wrong predictions (*False alarms* ($b$) and *Misses* ($c$)), all of which are divided by 2.

$$\text{Prediction skill index (PSI)} = \frac{\left(\frac{\frac{a}{n}-\left[\left(\frac{a+b}{n}\right)\left(\frac{a+c}{n}\right)\right]}{\sqrt{\left(\frac{a+b}{n}\right)\left(\frac{a+c}{n}\right)}} + \frac{\frac{d}{n}-\left[\left(\frac{b+d}{n}\right)\left(\frac{c+d}{n}\right)\right]}{\sqrt{\left(\frac{b+d}{n}\right)\left(\frac{c+d}{n}\right)}}\right) - \left(\frac{\frac{b}{n}-\left[\left(\frac{a+b}{n}\right)\left(\frac{b+d}{n}\right)\right]}{\sqrt{\left(\frac{a+b}{n}\right)\left(\frac{b+d}{n}\right)}} + \frac{\frac{c}{n}-\left[\left(\frac{a+c}{n}\right)\left(\frac{c+d}{n}\right)\right]}{\sqrt{\left(\frac{a+c}{n}\right)\left(\frac{c+d}{n}\right)}}\right)}{2}$$

$$-1 \leq \text{PSI} \leq 1$$

Our PSI score is bounded between -1 and 1. We set our reference score for no skill in prediction to be 0. The score of 1 means the perfect prediction skill, while the score of -1 means the worst prediction skill. Although the PSI score of -1 represents the poorest prediction skill performance, the information, however, can be valuable for the users because whatever the predictor predicts, the users could benefit by predicting the opposite.

We would like to clarify that, regardless of the methods used in prediction, the way in which the PSI method evaluates prediction skill is as follows. In every period (i.e. day, week, month, quarter, or year), a predictor, either human or machine, has to predict whether an event of interest will or will not occur in the next period. When the next period arrives, we compare the predicted outcome with the observed event, record if it is a *Hit* ($a$)*, False alarm* ($b$)*, Miss* ($c$)*, or Correct rejection* ($d$)*,* and continue to do this for *n* periods, where *n* is a large number. After *n* periods, we count the recorded data on the number of *Hits* ($a$)*, False alarms* ($b$)*, Misses* ($c$)*,* and *Correct rejections* ($d$) and use them to calculate the skill score for the PSI.

To illustrate our method, we present twenty-four numerical examples, using a total sample size (*n*) of 400 with different cell counts for *Hits* ($a$), *False alarms* ($b$), *Misses* ($c$), and *Correct rejections* ($d$) in order to show that our method not only satisfies many desirable properties that a prediction skill measure should have as discussed in Hogan and Mason (2012), but also could assess prediction skill under various situations.[1] Note that the total sample size and the number of different cell counts for the numerical examples are chosen arbitrarily. In practice, the total sample size and the cell counts could be any number taken from the actual predicted and observed events. We also show how the PSI method works in practice by using data on the Bank of Thailand's forecast of the growth of real gross domestic product (GDP) and the actual real GDP growth of Thailand between 2000 and 2019 as a case study.

In addition, we compare our PSI scores to those computed using five selected conventional methods for deterministic prediction skill evaluation as discussed in Wilks (2011). They are GSS, HSS, PSS, CSS, and ORSS. The definitions of these methods for deterministic prediction skill verification are provided in Table 6. The reason we choose these skill scores is that these methods for prediction skill verification are still widely used, as discussed in Section 1.

---

[1] The properties of the PSI along with those of five selected conventional methods for deterministic prediction skill verification are provided in Table A1 in the Appendix.

Table 6: Definitions of Selected Conventional Methods for Deterministic Prediction Skill Verification and Definition of Method for Deterministic Prediction Accuracy Verification.

| Name | Definition | Range |
|---|---|---|
| Gilbert skill score (GSS) | $\dfrac{a - \left[\dfrac{(a+b)(a+c)}{n}\right]}{a+b+c - \left[\dfrac{(a+b)(a+c)}{n}\right]}$ | [-⅓, 1] |
| Heidke skill score (HSS) | $\dfrac{a + d - \left[\dfrac{(a+b)(a+c)}{n}\right] - \left[\dfrac{(b+d)(c+d)}{n}\right]}{n - \left(\dfrac{(a+b)(a+c)}{n}\right) - \left[\dfrac{(b+d)(c+d)}{n}\right]}$ | [-1, 1] |
| Peirce skill score (PSS) | $\dfrac{ad - bc}{(b+d)(a+c)}$ | [-1, 1] |
| Clayton skill score (CSS) | $\dfrac{a}{a+b} - \dfrac{c}{c+d}$ | [-1, 1] |
| Odds ratio skill score (ORSS) | $\dfrac{ad - bc}{ad + bc}$ | [-1, 1] |
| Critical success index (CSI) | $\dfrac{a}{a+b+c}$ | [0, 1] |

Source: Hogan and Mason (2012).

Before we proceed, it is very important to clarify that prediction skill is not the same as prediction accuracy. Recall that prediction skill is referred to the relative accuracy of a set of forecasts with respect to some set of reference forecasts (Wilks, 2011). A highly skilled predictor, either human or machine, would generally tend to have a high rate of prediction accuracy, but the reverse may not be true. It depends upon the types of events/problems being predicted, their levels of difficulty, and the state of scientific knowledge used in prediction. For example, being able to predict today that tomorrow the sun will rise in the east and set in the west would probably hit a one-hundred percent prediction accuracy rate in our lifetime, but it hardly requires any skill. In contrast, predicting today whether tomorrow there will or will not be an economic crisis correctly both in terms of *Hits* ($a$) and *Correct rejections* ($d$) with no *False alarms* ($b$) and *Misses* ($c$) would require exceptional skill. To illustrate that prediction skill is not the same as prediction accuracy, we calculate the accuracy of the prediction using the critical success index (CSI) which is one of the most well-known methods for prediction accuracy verification. This issue is separately illustrated and discussed in Section 4. The definition of CSI is also provided in Table 6.

## 3. Results and Discussions

Before presenting and discussing the results, it is very important to reiterate that, irrespective of the techniques used in forecasting, the way in which the PSI method assesses the prediction skill is that, in every period (i.e. day, week, month, quarter, or year), a predictor, either human or machine, has to predict whether an event of interest will or will not occur in the next period. When the next period arrives, we compare the predicted outcome to the observed event, record if it is a *Hit* ($a$), *False alarm* ($b$), *Miss* ($c$), or *Correct rejection* ($d$), and continue to do this for $n$ periods, where $n$ is a large number. After $n$ periods, we count the recorded data on the number of *Hits* ($a$), *False alarms* ($b$), *Misses* ($c$), and *Correct rejections* ($d$) and use them to compute the PSI score. Note that the forecaster is not asked to predict today on what exact date(s) in the next $n$ days the event of interest will occur.

### *3.1 A case study using twenty-four numerical examples*

Skill scores based on our PSI method are reported in the sixth column of Table 7. Examples no. 1-4 show our reference score in that they do not require any skill to constantly predict the same value for *Hits* ($a$) or *False alarms* ($b$) or *Misses* ($c$) or *Correct rejections* ($d$), or to predict equal counts for each cell in the 2x2 contingency table as illustrated in example no. 5, or to constantly predict "*Yes*" as shown in example no. 6. In all cases, our PSI score is equal to 0. We also conduct a test for thirty random predictions and find that the average PSI score is more or less around 0.[2] The results are reported in Table A2 in the Appendix.

Examples no. 7-14 could be thought of as predicting any binary random event whose possible outcomes (i.e. yes/no, head/tail, up/down, or 0/1) in every period have a 50:50 chance of happening. We use examples no. 7-14 to show the range of the PSI scores from the best to the worst.[3] As shown in example no. 7, having roughly equal cell counts for both *Hits* ($a$) and *Correct rejections* ($d$) and no *False alarms* ($b$) and *Misses* ($c$) receives a perfect PSI score of 1. An example would be predicting correctly today that the outcome of a fair coin-toss tomorrow for 400 days in a roll, provided the probability of landing on head or tail on each day is 50:50. This is considered to be one of the most difficult tasks and requires extraordinary skills since random events, by definition, are extremely difficult to forecast given the current state of scientific knowledge. Note again that this should not be confused with binary events whose possible outcomes in every period have a 50:50 chance of occurring but are highly predictable, at least in our lifetime, and hence not difficult to forecast, such as the pattern of day and night cycle as discussed in Section 1. The higher the cell counts for *Hits* ($a$) and *Correct rejections* ($d$) in roughly equal proportions, the higher the PSI score. This is illustrated in example no. 8 where there are 175 *Hits* ($a$), 175 *Correct rejections* ($d$), 25 *False alarms* ($b$), and 25 *Misses* ($c$). In this case, the PSI score is equal to 0.750. In contrast, as shown in example no. 14, having equal cell counts for *False alarms* ($b$) and *Misses* ($c$) and no *Hits* ($a$) and *Correct rejections* ($d$) obtains the worst PSI score of -1. The higher the cell counts for *False alarms* ($b$) and *Misses* ($c$) in more or less similar proportions, the lower the PSI score. As illustrated in example no. 13, the cell counts for *Hits* ($a$), *Correct rejections* ($d$), *False alarms* ($b$), and *Misses* ($c$) are 25, 25, 175, and 175, respectively. Our PSI score, based on this example, is equal to -0.750.

---

[2] Given that the PSI method awards random predictions and always predicting the same value with the same score of 0, it satisfies the property of *equitability* in that it provides a no-skill benchmark against which a predictor can be said to have some measure of skill (Hogan & Mason, 2012).

[3] Since the PSI score ranges between -1 and 1, it satisfies the property of *boundedness*.

*Thailand and The World Economy | Vol. 40, No.3, September - December 2022*     | 41Table 7: Numerical Examples for Prediction Skill Verification Based on the PSI Method and the Other Conventional Skill Verification Methods.

| Example no. | Cell counts in 2x2 contingency table | | | | Skill verification methods | | | | | | Accuracy verification method |
|---|---|---|---|---|---|---|---|---|---|---|---|
| | Predicted/Observed events | | | | PSI | GSS | HSS | PSS | CSS | ORSS | CSI |
| | Yes/Yes (a) | Yes/No (b) | No/Yes (c) | No/No (d) | [-1, 1] | [-⅓, 1] | [-1, 1] | [-1, 1] | [-1, 1] | [-1, 1] | [0, 1] |
| 1 | 400 | 0 | 0 | 0 | 0.000 | 0.000 | 0.000 | 0.000 | 0.000 | 0.000 | 1.000 |
| 2 | 0 | 0 | 0 | 400 | 0.000 | 0.000 | 0.000 | 0.000 | 0.000 | 0.000 | N/A |
| 3 | 0 | 0 | 400 | 0 | 0.000 | 0.000 | 0.000 | 0.000 | 0.000 | 0.000 | 0.000 |
| 4 | 0 | 400 | 0 | 0 | 0.000 | 0.000 | 0.000 | 0.000 | 0.000 | 0.000 | 0.000 |
| 5 | 100 | 100 | 100 | 100 | 0.000 | 0.000 | 0.000 | 0.000 | 0.000 | 0.000 | 0.333 |
| 6 | 200 | 200 | 0 | 0 | 0.000 | 0.000 | 0.000 | 0.000 | 0.000 | 0.000 | 0.500 |
| 7 | 207 | 0 | 0 | 193 | 1.000 | 1.000 | 1.000 | 1.000 | 1.000 | 1.000 | 1.000 |
| 8 | 175 | 25 | 25 | 175 | 0.750 | 0.600 | 0.750 | 0.750 | 0.750 | 0.960 | 0.778 |
| 9 | 150 | 50 | 50 | 150 | 0.500 | 0.333 | 0.500 | 0.500 | 0.500 | 0.800 | 0.600 |
| 10 | 125 | 75 | 75 | 125 | 0.250 | 0.143 | 0.250 | 0.250 | 0.250 | 0.471 | 0.455 |
| 11 | 75 | 125 | 125 | 75 | -0.250 | -0.111 | -0.250 | -0.250 | -0.250 | -0.471 | 0.231 |
| 12 | 50 | 150 | 150 | 50 | -0.500 | -0.200 | -0.500 | -0.500 | -0.500 | -0.800 | 0.143 |
| 13 | 25 | 175 | 175 | 25 | -0.750 | -0.273 | -0.750 | -0.750 | -0.750 | -0.960 | 0.067 |
| 14 | 0 | 200 | 200 | 0 | -1.000 | -0.333 | -1.000 | -1.000 | -1.000 | -1.000 | 0.000 |
| 15 | 275 | 50 | 50 | 25 | 0.160 | 0.099 | 0.179 | 0.179 | 0.179 | 0.467 | 0.733 |
| 16 | 25 | 50 | 50 | 275 | 0.160 | 0.099 | 0.179 | 0.179 | 0.179 | 0.467 | 0.200 |
| 17 | 175 | 25 | 100 | 100 | 0.397 | 0.231 | 0.375 | 0.436 | 0.375 | 0.750 | 0.583 |
| 18 | 100 | 25 | 100 | 175 | 0.397 | 0.231 | 0.375 | 0.375 | 0.436 | 0.750 | 0.444 |
| 19 | 175 | 100 | 25 | 100 | 0.397 | 0.231 | 0.375 | 0.375 | 0.436 | 0.750 | 0.583 |
| 20 | 1 | 2 | 0 | 397 | 0.327 | 0.332 | 0.498 | 0.995 | 0.333 | 1.000 | 0.333 |
| 21 | 397 | 0 | 2 | 1 | 0.327 | 0.332 | 0.498 | 0.995 | 0.333 | 1.000 | 0.995 |
| 22 | 399 | 0 | 0 | 1 | 0.550 | 1.000 | 1.000 | 1.000 | 1.000 | 1.000 | 1.000 |
| 23 | 1 | 0 | 0 | 399 | 0.550 | 1.000 | 1.000 | 1.000 | 1.000 | 1.000 | 1.000 |
| 24 | 193 | 0 | 0 | 207 | 1.000 | 1.000 | 1.000 | 1.000 | 1.000 | 1.000 | 1.000 |

Source: Authors' calculation.

Notes: *Yes/Yes = Hits* ($a$), *Yes/No = False alarms* ($b$), *No/Yes = Misses* ($c$), and *No/No = Correct rejections* ($d$). Numerical examples from prediction accuracy verification using the CSI are also shown in order to illustrate the difference between prediction *skill* and prediction *accuracy*. Examples no. 1-6 show no prediction skill with a reference score = 0. Examples no. 7-14 show skill scores in predicting any binary random event where each of the possible outcomes has a 50:50 chance of happening. Examples no. 15-18 illustrate that swapping *Hits* ($a$) and *Correct rejections* ($d$) yields the same scores for PSI, GSS, HSS, and ORSS but it is not always true for PSS and CSS. Examples no. 17 and 19 show that swapping *False alarms* ($b$) and *Misses* ($c$) yields the same scores for PSI, GSS, HSS, and ORSS but this is not always true for PSS and CSS. Examples no. 18 and 19 illustrate that each method awards equal scores when swapping *Hits* ($a$) and *Correct rejections* ($d$) as well as swapping *False alarms* ($b$) and *Misses* ($c$). Examples no. 20-23 illustrate the ability of the PSI method and the other five conventional methods to produce nontrivial scores for predicting rare or extreme events that are difficult to forecast. Examples no. 20, 21, 22, 23, and 24 illustrate that the PSI method could distinguish between the degrees of difficulty in correct prediction of rare or extreme events commonly accepted as very difficult to forecast and random events by awarding different scores while the other five conventional methods do not make this distinction among examples no. 22, 23 (an economic crisis), and 24 (a fair coin-toss) and award the same score. Examples no. 1, 5, and 6 clearly show the distinction between prediction *skill* and prediction *accuracy*. In addition, the difference in scores between examples no. 6 and 7 is 1 for PSI while that for CSI is only 0.5. Examples no. 21 and 22 show a similar result. The difference in scores between these two examples is 0.223 for PSI, whereas that for CSI is merely 0.005.

We would like to note that our method treats *Hits* ($a$), *False alarms* ($b$), *Misses* ($c$), and *Correct rejections* ($d$) equally. This can be illustrated by examples no. 15-18 that swapping the cell counts for *Hits* ($a$) and those for *Correct rejections* ($d$) yields the same PSI scores. In case of *False alarms* ($b$) and *Misses* ($c$), examples no. 17 and 19 show that swapping the cell counts for *False alarms* ($b$) and those for *Misses* ($c$) also produces the same PSI score.[4] Note that, in the circumstances where *Misses* ($c$) are more important than *False alarms* ($b$), more weight could be assigned to *Misses* ($c$) relative to *False alarms* ($b$) and vice versa. Moreover, our method awards an equal score for swapping the cell counts for *Hits* ($a$) and those for *Correct rejections* ($d$) as well as swapping the cell counts for *False alarms* ($b$) and those for *Misses* ($c$) as illustrated by examples no. 18 and 19.[5]

In addition, our method could evaluate prediction skill for rare or extreme events that are considered difficult to forecast reasonably well as shown in examples no. 20 and 21. There is only one cell count for *Hits* ($a$) in example no. 20 but note that *"Yes"* was predicted three times, and the event did occur once out of 400 predictions. Our PSI score for this example is 0.327. As earlier discussed in Section 1, financial and economic crises, the success of start-up unicorns, droughts, floods, wildfires, earthquakes, wars, and pandemics are rare or extreme events commonly accepted as very difficult to forecast. For the purpose of illustrating our method, an event for example no. 20 is forecasting today that tomorrow there will be an economic crisis correctly once out of 400 consecutive days with 2 *False alarms* ($b$), no *Misses* ($c$), and 397 *Correct rejections* ($d$). Recall that this should not be confused with events that are commonly considered as rare or extreme but not difficult to forecast given the current state of scientific knowledge like the solar eclipse also discussed in Section 1. Example no. 21 shows the opposite case, where there is only one cell count for *Correct rejection* ($d$), but note again that "*No*" was predicted three times out of 400 predictions and there is one time that the event was correctly predicted. Our PSI method awards a score of 0.327 for this example. Examples no. 20 and 21 also show that swapping the cell counts for *Hits* ($a$) and those for *Correct rejections* ($d$) as well as swapping the cell counts for *False alarms* ($b$) and those for *Misses* ($c$) yield an equal score. This suggests that a relabelling of events and non-events has no effect on the ability of our method to verify prediction skill of rare or extreme events as noted by Hogan and Mason (2012).

Furthermore, our method would be able to distinguish between different degrees of difficulty in the correct prediction of rare or extreme events that are difficult to forecast and random events that are extremely more difficult to forecast given the present state of scientific knowledge. This could be illustrated using examples no. 20, 21, 22, 23, and 24.[6]

---

[4] Examples no. 17 and 19 suggest that the PSI method has the property of *transpose symmetry* since swapping of *False alarms* ($b$) and *Misses* ($c$) yields the same score. This ensures that *False alarms* ($b$) and *Misses* ($c$) are treated equally.

[5] Examples no. 18 and 19 indicate that the PSI method has the property of *complement symmetry* because it awards an equal score for swapping *Hits* ($a$) and *Correct rejections* ($d$) as well as swapping *False alarms* ($b$) and *Misses* ($c$). According to Hogan and Mason (2012), *complement symmetry* is a useful property for verification of rare or extreme events, where it is a much higher priority to get the occurrences right than the non-occurrences.

[6] Note that if we slightly alter the number of cell count for *Hits* ($a$) in example no. 21 from 397 to 399 while the number of other cell counts remains unchanged and call it example no. $21^*$, it can be shown that the PSI method has the property of *non-regularity*. Compared with example no. 22, example no. $21^*$ under-predicts the occurrence, but all forecasts issued are correct. That is there are misses but no false alarms (bias score = 0.995). A regular measure would award

In all five cases, although there are almost *Hits* ($a$) and/or *Correct rejections* ($d$) and hardly any *False alarms* ($b$) and *Misses* ($c$), example no. 24 is considered to be significantly more difficult to forecast and requires enormously higher skills than examples no. 20, 21, 22 and 23. Our PSI method, therefore, awards a score of 0.327 for examples no. 20 and 21, a score of 0.550 for examples no. 22 and 23, and a perfect score of 1 for example no. 24. Similar to example no. 7, example no. 24 could be thought of as correctly predicting today the outcome of a fair coin-toss (a random event) tomorrow for 400 consecutive days with no *False alarms* ($b$) and *Misses* ($c$) provided that the probability of landing on head or tail on each day is 50:50, while examples no. 20 and 23 could be thought of as correctly predicting today that there will or will not be an economic crisis (a rare or extreme event commonly accepted as very difficult to forecast) tomorrow correctly both in terms of *Hits* ($a$) and *Correct rejections* ($d$) with no or a couple of *False alarms* ($b$) and *Misses* ($c$).[7] Our PSI method shows that, for 400 days in a roll, correctly predicting today the outcome of a fair coin-toss (a random event) tomorrow given that its chance of landing on head or tail on each day is 50:50, *is extremely much harder* than correctly predicting today whether there will be an economic crisis (a rare or extreme event commonly accepted as very difficult to forecast) tomorrow. Therefore, our PSI method awards example no. 23 a score of 0.550 and example no. 24 a score of 1. Examples no. 23 and 24 clearly indicates that, *when evaluating the prediction skill, perfect forecasts do not always have a perfect score of 1 as conventionally perceived*. It depends upon the types of events/problems being forecasted, their levels of difficulty, and the state of scientific knowledge used in forecasting.

     Next, we compare skill scores calculated using our PSI method with skill scores computed using five selected conventional methods for prediction skill evaluation, namely, GSS, HSS, PSS, CSS, and ORSS. It can be seen from Table 7 that there are no differences among these methods in case of no prediction skill (examples no. 1-6). All methods including ours award a score of 0. In addition, ranging from the highest to the lowest, as illustrated in examples no. 7-14, our PSI scores are identical to those of HSS, PSS, and CSS. While swapping the cell counts for *Hits* ($a$) and those for *Correct rejections* ($d$), as shown in examples no. 15-18, yields the same scores based on four methods, including ours, namely, PSI, GSS, HSS, and ORSS, our numerical examples suggest that this is not always true for PSS and CSS (examples no. 17 and 18). Likewise, swapping the cell counts for *False alarms* ($b$) and those for *Misses* ($c$), as shown in examples no. 17 and 19, produces the same scores in four methods, including ours, namely, PSI, GSS, HSS, and ORSS. Again, numerical examples no. 17 and 19 indicate that this is not always true for PSS and CSS.

     In the case of evaluating prediction skills for rare or extreme events that are difficult to forecast, as shown in examples no. 20 and 21, our PSI score (0.327) is slightly

---

example no. 21[*] the same score as example no. 22, a perfect unbiased forecaster, that issues no misses or false alarms (bias score = 1). Example no. 21[*] also shows that the PSI method *does not award a biased forecast a perfect score of 1*.

[7] Examples no. 20 and 23 show that the PSI method satisfies the property of *difficulty to hedge* since a predictor is not encouraged to predict rare or extreme events that are difficult to forecast on the basis of their low probability alone. This could be illustrated by example no. 2 where always predicting "*No*" for rare or extreme events would receive a score of 0 based on the PSI method. In addition, examples no. 20 and 23 demonstrate that the PSI method satisfies the property of *non-degeneracy for rare or extreme events* that are difficult to forecast since both examples show that rare or extreme events can be skillfully forecasted by awarding nontrivial scores. By having the property of *non-degeneracy for rare or extreme events*, the PSI method, therefore, has the property of *non-linearity*.

lower than those of GSS (0.332) and CSS (0.333). PSS produces almost a perfect score, which is 0.995, while HSS awards a score of 0.498. ORSS is the only method that awards a perfect score of 1 for correctly predicting rare or extreme events that are difficult to forecast. Whereas six different methods yield different skill scores ranging from 0.327 to 1 as shown in examples No. 20 and 21, each method awards equal scores when swapping the cell counts for *Hits* ($a$) and those for *Correct rejections* ($d$) as well as swapping the cell counts for *False alarms* ($b$) and those for *Misses* ($c$).

Furthermore, our PSI method awards different scores for correctly predicting rare or extreme events commonly accepted as difficult to forecast and random events depending upon the levels of difficulty in predicting such events as shown in examples no. 20, 21, 22, 23, and 24. However, GSS, HSS, PSS, CSS, and ORSS *do not* make this distinction among examples no. 22, 23, and 24 and award all three events the same score of 1. These results clearly indicate that GSS, HSS, PSS, CSS, and ORSS treat the degree of difficulty in correct prediction of an economic crisis (a rare or extreme event commonly accepted as very difficult to forecast) both in terms of *Hits* ($a$) and *Correct rejections* ($d$) with no *False alarms* ($b$) and *Misses* ($c$) *(*example no. 23) identical to correct prediction of a fair coin-toss (a random event generally accepted as extremely difficult to forecast based on the current state of scientific knowledge) with no *False alarms* ($b$) and *Misses* ($c$) (example no. 24) by awarding the same skill score of 1, whereas our PSI method treats the two events quite differently by awarding the former a skill score of 0.550 and a skill score of 1 for the latter. This indicates that perfect forecasts do not always receive a perfect score of 1, as shown in Figure 1. Recall that the way in which the PSI evaluates the prediction skill is that, in the case of a fair coin-toss (a random event), in every day, the predictor has to predict today the outcome of a fair coin-toss tomorrow for 400 consecutive days, provided on each day, there is a 50:50 chance of landing on head or tail. For the prediction of an economic crisis (a rare or extreme event), the predictor is asked every day for 400 consecutive days whether tomorrow there will or will not be an economic crisis. It is very important to note that the predictor is not asked to predict today on what exact date(s) in the next 400 days an economic crisis will occur.

While choosing among skill scores may vary from situation to situation (Gandin & Murphy, 1992), one reasonable choice is that the chosen skill score should award the same score for random prediction and constantly predicting the same value (*equitability*). In addition, correct predictions of rare or extreme events that are difficult to forecast should be weighed more strongly than correct predictions of more common events (*difficulty to hedge*) which should discourage distortion of prediction toward the more common event in order to artificially inflate the skill score (Wilks, 2011). Viewed this way, along with other desirable properties of skill verification measures, namely, *boundedness*, *non-degeneracy for rare or extreme events*, *non-linearity*, *non-regularity*, *imperfect score for biased forecast*, *transpose symmetry*, and *complement symmetry*, the PSI method should pass the required tests for a robust performance measure and could be used as an alternative method to assess skill in forecasting deterministic binary events.[8]

---

[8] It should be noted that although the PSI method satisfies many desirable properties of a forecast verification measure, it is not *base-rate independent*. That is a change in base rate $\left(\frac{Hits\ (a)\ +\ Misses\ (c)}{n}\right)$ could affect the performance of the PSI. For further discussion on the property of *base-rate independence*, see Ferro and Stephenson (2011).

Figure 1: A comparison of skill scores between the perfect forecast of a rare or extreme event commonly accepted as very difficult to forecast (i.e. an economic crisis) and that of a random event generally accepted as extremely difficult to forecast based on the current state of scientific knowledge (i.e. a fair coin-toss) with the number of forecast periods ($n$) = 400.

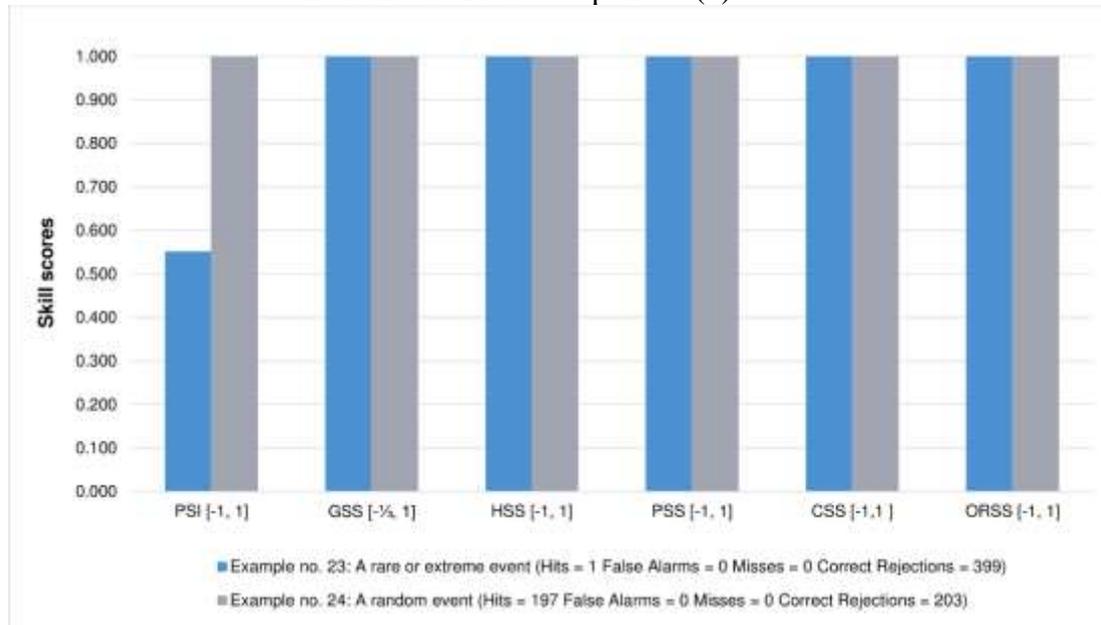

Source: Authors' calculation.
Note: Given that the perfect forecast of a random event (example no. 24) is extremely much harder than that of a rare or extreme event (example no. 23), the two forecasts are not awarded the same perfect score of 1 as indicated by the PSI, whereas the other five conventional prediction skill verification methods do not make this distinction and award both forecasts the same perfect score of 1.

### *3.2 A case study using the real GDP growth forecast*

The GDP growth forecast is one of the most followed figures around the world. Before using the GDP growth forecast for any decision-making purpose, it is necessary to know whether a forecaster, either human or machine, is skillful enough to predict the direction of future GDP growth. While the magnitude of GDP growth does matter, we view that a priority must be given to the direction. Once we get the direction of GDP growth correct, its magnitude can later be adjusted. To illustrate the use of the PSI method in assessing skill in predicting the direction of GDP growth in practice, this study uses the Bank of Thailand's real GDP growth forecast between 2000 and 2019 (total of 78 quarters) as a case study. As described in Section 2, we compare the annual real GDP growth forecast made in each quarter by the Bank of Thailand with the actual real GDP growth of Thailand in that year (see Table A3 in the Appendix), record the predicted outcome in each quarter, whether it is a *Hits (Up/Up)*, *False alarms (Up/Down)*, *Misses (Down/Up)*, or *Correct rejections (Down/Down)*, and use all the recorded outcomes to calculate the PSI score. Table 8 reports our results.

Table 8: The Evaluation of Prediction Skill of the Bank of Thailand's Real GDP Growth Forecast between 2000 and 2019 (Total of 78 Quarters) Using the PSI Method and Other Skill Verification Methods.

| Cell counts in 2x2 contingency table | | | | Skill verification methods | | | | | | Accuracy verification method |
|---|---|---|---|---|---|---|---|---|---|---|
| Predicted/Observed events | | | | PSI | GSS | HSS | PSS | CSS | ORSS | CSI |
| Up/Up (a) | Up/Down (b) | Down/Up (c) | Down/Down (d) | [-1, 1] | [-⅓, 1] | [-1, 1] | [-1, 1] | [-1, 1] | [-1, 1] | [0, 1] |
| 34 | 4 | 4 | 36 | 0.795 | 0.659 | 0.795 | 0.795 | 0.795 | 0.974 | 0.810 |

Source: Authors' calculation.
Notes: Up/Up = *Hits* ($a$), Up/Down = *False alarms* ($b$), Down/Up = *Misses* ($c$), and Down/Down = *Correct rejections* ($d$). The result from prediction accuracy verification using the CSI is also reported in order to illustrate the difference between prediction *skill* and prediction *accuracy*.

In terms of the movement of the real GDP growth (*Up* vs. *Down*), the Bank of Thailand correctly predicts the direction of the real GDP growth 70 out of 78 predictions, with 34 *Hits (Up/Up)*, 4 *False alarms (Up/Down)*, 4 *Misses (Down/Up)*, and 36 *Correct rejections (Down/Down)*. While *Misses* ($c$) are usually considered as more important than *False alarms* ($b$) in many scientific professions such as atmospheric science, medicine, and seismology, the real GDP growth forecast is one example where *Misses (Down/Up)* may have relatively fewer negative impacts than *False alarms (Up/Down)*. Given these prediction outcomes, our PSI method awards a score of 0.795. Note that the other five conventional skill verification methods also award scores in the same fashion. HSS, PSS, and CSS give the same score of 0.795, while GSS awards a score of 0.659 and ORSS awards a score of 0.974. These skill scores indicate that the performance of the Bank of Thailand in forecasting the direction of real GDP growth is better than a fair coin-toss. While the overall performance of the Bank of Thailand is better than chance, it should be noted that it failed to predict the direction of the Thai real GDP growth in 2008 when the Thai economy was negatively affected by the global financial crisis which is considered a rare or extreme event. This is not surprising given that both macro and micro uncertainty appear to increase significantly during the economic downturns (Bloom, 2014).

## 4. Difference between Prediction Skill and Prediction Accuracy

Prediction skill is often misunderstood or confused with prediction accuracy. As previously noted, prediction skill is defined as the relative accuracy of a set of forecasts with respect to some set of reference forecasts (Wilks, 2011). Generally, predictors, either human or machine, with higher skill would probably achieve higher prediction accuracy than those with lower skill. However, the reverse is not necessarily true. We use examples no. 1, 5, 6 and 7 as well as examples no. 21 and 22 as shown in Table 7 to illustrate the difference between prediction skill and prediction accuracy.

Examples no. 1, 5, and 6 show that it is possible to achieve various degrees of prediction accuracy without having any prediction skill. In example no. 1, CSI, which is a well-known method for verifying prediction accuracy, gives a perfect score of 1 for predicting *Hits* ($a$) 400 times, but all prediction skill verification methods, including ours, award a no-skill score of 0. Examples no. 5 and 6 also show similar results. As shown in example no. 5, CSI gives a score of 0.333 for predicting equal counts for *Hits* ($a$), *False alarms* ($b$), *Misses* ($c$), and *Correct rejections* ($d$) whereas all skill scores, including ours, award a no-skill score of 0. All prediction skill verification methods, including ours, also award a no-skill score of 0 for constantly predicting *"Yes"* as illustrated in example no. 6 while CSI awards a score of 0.500.

Focusing particularly on the PSI and CSI, the difference between the two could be illustrated using examples no. 6, 7, 21 and 22. PSI awards a no-skill score of 0 for example no. 6 and a perfect score of 1 for example no. 7 (a random event whose binary outcomes in each period have a 50:50 chance of happening). CSI also awards a perfect score of 1 for example no. 7 but a score of 0.500 for example no. 6. Thus, the difference in scores between example no. 6 (no skill by always predicting *"Yes"*) and example no. 7 (exceptional skill by correctly predicting *Hits* ($a$) and *Correct rejections* ($d$) in roughly equal proportions with no *False alarms* ($b$) and *Misses* ($c$)) is 1 for the PSI, while that for CSI is only 0.5. Examples no. 21 and 22 also show that the PSI awards quite different scores (0.327 vs. 0.550) for different levels of difficulty in correct prediction of rare or extreme events commonly accepted as difficult to forecast, while CSI awards somewhat similar scores (0.995 vs. 1).

These examples illustrate that prediction skill *is not* the same as prediction accuracy. The main difference is that prediction skill verification methods take the degrees of difficulty in correctly predicting the events into account while CSI does not. As clearly shown in example no. 1, it is possible to have a one-hundred percent prediction accuracy rate with no prediction skill. Usually, we would expect the scores of prediction skill and prediction accuracy to go in tandem as for the case of the real GDP growth forecast of the Bank of Thailand as reported in Table 8 where the PSI awards a *skill* score of 0.795 and CSI awards an *accuracy* score of 0.810.

## 5. Conclusions and Remarks

This study devises the prediction skill index (PSI) as an alternative method for evaluating skill in deterministic forecasts of binary events. Using twenty-four numerical examples, we show that our method not only satisfies many desirable properties of forecast verification measures but also could evaluate skill in deterministic forecast of binary events under a variety of circumstances. In addition, we compare our skill scores with those computed using five selected conventional prediction skill verification measures, namely, GSS, HSS, PSS, CSS, and ORSS. The results indicate that, in most cases, our PSI method awards skill scores not much different from those computed by using HSS. The key difference between the PSI and HSS as well as the other four conventional skill scores, however, is the ability to separate the difference between the perfect forecast of rare or extreme events that are commonly accepted as very difficult to forecast and that of random events (i.e. predicting correctly today that there will or will not be an economic crisis (a rare or extreme event commonly accepted as very difficult to forecast) tomorrow both in terms of *Hits* ($a$) and *Correct rejections* ($d$) with no *False alarms* ($b$) and *Misses* ($c$) (example no. 23) vs. correctly predicting today the tomorrow outcome of a fair coin-toss (a random event that are generally accepted as extremely difficult to forecast given the current state of scientific knowledge) with no *False alarms* ($b$) and *Misses* ($c$) provided the probability of landing on head or tail on each day is 50:50 (example no. 24)). This implies that a perfect forecast does not necessarily have a perfect score of 1. It depends upon the degree of difficulty of the event/problem being forecasted. We also show that prediction skill should not be confused with prediction accuracy since they are not the same. The numerical examples from our PSI and the other five prediction skill verification methods vs. those from CSI clearly indicate that it is possible to have a high rate of prediction accuracy, but no skill in prediction is required.

In addition to the numerical examples, this study uses the real GDP growth forecast of the Bank of Thailand between 2000 and 2019 as a case study in order to show how the PSI method can be used to evaluate the prediction skills of the forecaster in practice. While the Bank of Thailand failed to predict the downturn of the Thai economy adversely impacted by the global financial crisis in 2008, the overall results indicate that its real GDP growth forecast is fairly accurate and better than a fair coin-toss.

Our PSI method is less prone to be manipulated since it passes the required criteria for a robust performance measure in that it not only awards the same score for random prediction and constantly predicting the same value, but also yields nontrivial skill scores for correct prediction of rare or extreme events that are difficult to forecast. In addition, the PSI method treats all cell counts in the 2x2 contingency table, namely, *Hits* ($a$), *False alarms* ($b$), *Misses* ($c$), and *Correct rejections* ($d$) equally. This can be illustrated by swapping the cell counts for *Hits* ($a$) and those for *Correct rejections* ($d$) or swapping the cell counts for *False alarms* ($b$) and those for *Misses* ($c$) which would result in the same scores. It is worth to note again that more weight could be assigned to *Misses* ($c$) relative to *False alarms* ($b$) in case the consequences from *Misses* ($c$) are considered to have higher impacts than those from *False alarms* ($b$) and vice versa. Swapping the cell counts for *Hits* ($a$) and those for *Correct rejections* ($d$) as well as swapping the cell counts for *False alarms* ($b$) and those for *Misses* ($c$) also yield an equal score, suggesting that our method is invariant to a relabelling of events and non-events and could be used to assess the prediction skill of rare or extreme events that are difficult to forecast. Furthermore, our PSI method would be able to distinguish the difference between the perfect forecast of rare or extreme events, commonly accepted as very difficult to forecast and that of random events. If the random events and their consequences, either good or bad, do not present a serious challenge or concern, then the existing conventional prediction skill verification methods could be employed. Otherwise, knowing whether or not a forecaster, either human or machine, is skillful enough to predict future random events could be useful for both theoretical and practical purposes.

Given the increased reliance on predictions in various scientific disciplines involving decision-making and the impacts of rare or extreme events and random events on human societies and our planet, we hope that both experts and general audiences, irrespective of their disciplines and walks of life, will find our PSI method applicable and will use it in combination with other measures of prediction quality in order to evaluate the overall performance of deterministic prediction of binary events for their scientific, economic, and administrative purposes (Woodcock, 1976).

# Acknowledgements

This study is supported by Grants for Development of New Faculty Staff, Ratchadaphiseksomphot Endowment Fund, Chulalongkorn University, Contract No. DNS 61_041_26_001_1. We sincerely thank Dr. Suradit Holasut for guidance and comments.

# Appendix

Table A1: The Properties of the PSI Method and Those of Five Selected
Conventional Methods for Deterministic Prediction Skill Verification.

| Methods | Equitability | Difficulty to hedge | Non-degeneracy | Base-rate independence | Boundedness | Linearity | Regularity | Biased forecast can get perfect score | Transpose symmetry | Complement symmetry |
|---|---|---|---|---|---|---|---|---|---|---|
| PSI | ✓ | ✓ | ✓ | | ✓ | | | ✓ | ✓ | ✓ |
| GSS | ✓ | ✓ | | | ✓ | | | | ✓ | |
| HSS | ✓ | ✓ | | | ✓ | ✓ | | | ✓ | ✓ |
| PSS | ✓ | ✓ | | ✓ | ✓ | ✓ | | | | ✓ |
| CSS | ✓ | ✓ | | | ✓ | ✓ | | | | ✓ |
| ORSS | ✓ | ✓ | | ✓ | ✓ | | ✓ | ✓ | ✓ | ✓ |

Source: Adapted from Hogan and Mason (2012).

Table A2: The Evaluation of Thirty Random Predictions Based on the PSI Method and Other Skill Verification Methods.

| Random prediction no. | Elements in 2x2 contingency table | | | | | Skill verification methods | | | | | |
|---|---|---|---|---|---|---|---|---|---|---|---|
| | Predicted/Observed events | | | | | PSI | GSS | HSS | PSS | CSS | ORSS |
| | *Yes/Yes* (*a*) | *Yes/No* (*b*) | *No/Yes* (*c*) | *No/No* (*d*) | Total | [-1,1] | [-⅓,1] | [-1,1] | [-1,1] | [-1,1] | [-1,1] |
| 1 | 517 | 912 | 178 | 174 | 1781 | -0.111 | -0.039 | -0.081 | -0.096 | -0.144 | -0.287 |
| 2 | 85 | 845 | 212 | 31 | 1173 | -0.670 | -0.166 | -0.398 | -0.678 | -0.781 | -0.971 |
| 3 | 333 | 390 | 849 | 506 | 2078 | -0.157 | -0.067 | -0.145 | -0.154 | -0.166 | -0.325 |
| 4 | 234 | 990 | 471 | 119 | 1814 | -0.571 | -0.198 | -0.494 | -0.561 | -0.607 | -0.887 |
| 5 | 554 | 811 | 682 | 743 | 2790 | -0.073 | -0.035 | -0.073 | -0.074 | -0.073 | -0.147 |
| 6 | 681 | 822 | 422 | 627 | 2552 | 0.050 | 0.025 | 0.048 | 0.050 | 0.051 | 0.104 |
| 7 | 872 | 237 | 589 | 914 | 2612 | 0.391 | 0.234 | 0.379 | 0.391 | 0.394 | 0.702 |
| 8 | 359 | 554 | 125 | 726 | 1764 | 0.268 | 0.138 | 0.242 | 0.309 | 0.246 | 0.580 |
| 9 | 687 | 19 | 492 | 766 | 1964 | 0.561 | 0.340 | 0.507 | 0.558 | 0.582 | 0.965 |
| 10 | 228 | 851 | 667 | 50 | 1796 | -0.701 | -0.256 | -0.689 | -0.690 | -0.719 | -0.961 |
| 11 | 702 | 339 | 97 | 238 | 1376 | 0.322 | 0.183 | 0.309 | 0.291 | 0.385 | 0.671 |
| 12 | 150 | 365 | 399 | 215 | 1129 | -0.357 | -0.151 | -0.357 | -0.356 | -0.359 | -0.637 |
| 13 | 457 | 786 | 106 | 806 | 2155 | 0.273 | 0.129 | 0.229 | 0.318 | 0.251 | 0.631 |
| 14 | 543 | 273 | 142 | 67 | 1025 | -0.011 | -0.006 | -0.011 | -0.010 | -0.014 | -0.032 |
| 15 | 91 | 393 | 754 | 816 | 2054 | -0.241 | -0.104 | -0.232 | -0.217 | -0.292 | -0.599 |
| 16 | 407 | 283 | 890 | 50 | 1630 | -0.414 | -0.138 | -0.320 | -0.536 | -0.357 | -0.850 |
| 17 | 488 | 324 | 634 | 820 | 2266 | 0.157 | 0.082 | 0.152 | 0.152 | 0.165 | 0.322 |
| 18 | 604 | 131 | 914 | 452 | 2101 | 0.156 | 0.065 | 0.123 | 0.173 | 0.153 | 0.390 |
| 19 | 7 | 787 | 109 | 747 | 1650 | -0.201 | -0.058 | -0.122 | -0.453 | -0.119 | -0.885 |
| 20 | 185 | 972 | 941 | 516 | 2614 | -0.485 | -0.196 | -0.487 | -0.489 | -0.486 | -0.811 |
| 21 | 325 | 996 | 459 | 249 | 2029 | -0.387 | -0.146 | -0.342 | -0.385 | -0.402 | -0.699 |
| 22 | 238 | 532 | 286 | 412 | 1468 | -0.104 | -0.047 | -0.099 | -0.109 | -0.101 | -0.216 |
| 23 | 907 | 192 | 178 | 361 | 1638 | 0.478 | 0.326 | 0.492 | 0.489 | 0.495 | 0.811 |
| 24 | 522 | 237 | 34 | 36 | 829 | 0.104 | 0.046 | 0.087 | 0.071 | 0.202 | 0.400 |
| 25 | 598 | 72 | 32 | 158 | 860 | 0.630 | 0.508 | 0.673 | 0.636 | 0.724 | 0.952 |
| 26 | 915 | 435 | 644 | 262 | 2256 | -0.034 | -0.017 | -0.034 | -0.037 | -0.033 | -0.078 |
| 27 | 587 | 169 | 239 | 599 | 1594 | 0.491 | 0.324 | 0.489 | 0.491 | 0.491 | 0.794 |
| 28 | 13 | 972 | 552 | 146 | 1683 | -0.797 | -0.263 | -0.715 | -0.846 | -0.778 | -0.993 |
| 29 | 747 | 675 | 123 | 951 | 2496 | 0.421 | 0.240 | 0.386 | 0.443 | 0.411 | 0.791 |
| 30 | 510 | 303 | 571 | 72 | 1456 | -0.286 | -0.120 | -0.273 | -0.336 | -0.261 | -0.650 |
| **Average skill score** | | | | | | **-0.043** | **0.021** | **-0.025** | **-0.055** | **-0.038** | **-0.064** |

Source: Authors' calculation.
Notes: Yes/Yes = *Hits* (*a*), Yes/No = *False alarms* (*b*), No/Yes = *Misses* (*c*), and No/No = *Correct rejections* (*d*). All random numbers used in this study were generated using =rand() formula in Microsoft Excel 2019. The results from all skill verification methods indicate that the average scores are more or less around 0.

Table A3: Data on the Growth of Real Gross Domestic Product (GDP) Forecast of the Bank of Thailand and the Actual Real GDP Growth of Thailand between 2000 and 2019.

| Year | Quarter | Real GDP growth forecast for the entire year made in each quarter (%) | Actual real GDP growth (% year-on-year) |
|---|---|---|---|
| 1999 | - | - | 4.4 |
| 2000 | Q1 | - | 4.8 |
|  | Q2 | - |  |
|  | Q3 | 5.0 |  |
|  | Q4 | 4.8 |  |
| 2001 | Q1 | 3.8 | 2.2 |
|  | Q2 | 3.3 |  |
|  | Q3 | 2.5 |  |
|  | Q4 | 1.6 |  |
| 2002 | Q1 | 2.5 | 5.3 |
|  | Q2 | 3.0 |  |
|  | Q3 | 3.5 |  |
|  | Q4 | 4.3 |  |
| 2003 | Q1 | 4.0 | 7.1 |
|  | Q2 | 4.0 |  |
|  | Q3 | 5.0 |  |
|  | Q4 | 6.0 |  |
| 2004 | Q1 | 6.8 | 6.3 |
|  | Q2 | 7.3 |  |
|  | Q3 | 6.5 |  |
|  | Q4 | 6.0 |  |
| 2005 | Q1 | 5.8 | 4.6 |
|  | Q2 | 5.0 |  |
|  | Q3 | 4.0 |  |
|  | Q4 | 4.5 |  |
| 2006 | Q1 | 5.3 | 5.1 |
|  | Q2 | 4.8 |  |
|  | Q3 | 4.5 |  |
|  | Q4 | 4.8 |  |
| 2007 | Q1 | 4.5 | 5.0 |
|  | Q2 | 4.3 |  |
|  | Q3 | 4.5 |  |
|  | Q4 | 4.6 |  |
| 2008 | Q1 | 5.3 | 2.5 |
|  | Q2 | 5.4 |  |
|  | Q3 | 5.3 |  |
|  | Q4 | 4.7 |  |
| 2009 | Q1 | 1.0 | -2.3 |
|  | Q2 | -2.5 |  |
|  | Q3 | -3.8 |  |
|  | Q4 | -3.0 |  |

| Year | Quarter | Real GDP growth forecast for the entire year made in each quarter (%) | Actual real GDP growth (% year-on-year) |
|---|---|---|---|
| 2010 | Q1 | 4.3 | 7.8 |
|  | Q2 | 5.1 |  |
|  | Q3 | 7.0 |  |
|  | Q4 | 7.7 |  |
| 2011 | Q1 | 4.0 | 0.1 |
|  | Q2 | 4.1 |  |
|  | Q3 | 4.1 |  |
|  | Q4 | 2.6 |  |
| 2012 | Q1 | 4.9 | 6.5 |
|  | Q2 | 6.0 |  |
|  | Q3 | 5.7 |  |
|  | Q4 | 5.7 |  |
| 2013 | Q1 | 4.9 | 2.9 |
|  | Q2 | 5.1 |  |
|  | Q3 | 4.2 |  |
|  | Q4 | 3.7 |  |
| 2014 | Q1 | 2.7 | 0.8 |
|  | Q2 | 1.5 |  |
|  | Q3 | 1.5 |  |
|  | Q4 | 0.8 |  |
| 2015 | Q1 | 3.8 | 2.9 |
|  | Q2 | 3.0 |  |
|  | Q3 | 2.7 |  |
|  | Q4 | 2.8 |  |
| 2016 | Q1 | 3.1 | 3.2 |
|  | Q2 | 3.1 |  |
|  | Q3 | 3.2 |  |
|  | Q4 | 3.2 |  |
| 2017 | Q1 | 3.4 | 3.9 |
|  | Q2 | 3.5 |  |
|  | Q3 | 3.8 |  |
|  | Q4 | 3.9 |  |
| 2018 | Q1 | 4.1 | 4.1 |
|  | Q2 | 4.4 |  |
|  | Q3 | 4.4 |  |
|  | Q4 | 4.2 |  |
| 2019 | Q1 | 3.8 | 2.4 |
|  | Q2 | 3.3 |  |
|  | Q3 | 2.8 |  |
|  | Q4 | 2.5 |  |

Source: Bank of Thailand (2020).
Note: The actual real GDP growth in 1999 is included in order to compare the direction of the actual real GDP growth with the forecasts made by the Bank of Thailand in 2000.